\documentclass[12pt]{article}

\pdfoutput=1

\usepackage{graphics}
\usepackage{amssymb}
\usepackage{amsmath}
\usepackage{tikz}

\newcommand{\beq}{\begin{equation}}

\newcommand{\eeq}{\end{equation}}
\newcommand{\bea}{\begin{eqnarray}}
\newcommand{\eea}{\end{eqnarray}}

\begin{document}

\begin{center}
${}$\\
\vspace{100pt}
{ \Large \bf How round is the quantum de Sitter universe? 
}

\vspace{46pt}

{\sl N. Klitgaard}
and {\sl R. Loll}

\vspace{24pt}
{\footnotesize

Institute for Mathematics, Astrophysics and Particle Physics, Radboud University \\ 
Heyendaalseweg 135, 6525 AJ Nijmegen, The Netherlands.\\ 
\vspace{12pt}
{email: n.klitgaard@science.ru.nl, r.loll@science.ru.nl}\\
}
\vspace{48pt}

\end{center}

%\addtolength{\baselineskip}{0.20\baselineskip}
\vspace{0.8cm}

\begin{center}
{\bf Abstract}
\end{center}

\noindent 
We investigate the quantum Ricci curvature, which was introduced in earlier work, in full, four-dimensional quantum gravity,
formulated nonperturbatively in terms of Causal Dynamical Triangulations (CDT). A key finding of the CDT approach is the emergence of
a universe of de Sitter-type, as evidenced by the successful 
matching of Monte Carlo measurements of the quantum dynamics of the global scale factor 
with a semiclassical minisuperspace model. 
An important question is whether the quantum universe exhibits semiclassicality also with regard to its more local
geo\-metric properties. 
Using the new quantum curvature observable, we examine whether the (quasi-)local properties of the quantum geometry 
resemble those of a constantly curved space.
We find evidence that on sufficiently large scales the curvature behaviour 
is compatible with that of a four-sphere, thus strengthening the interpretation of the dynamically 
generated quantum universe in terms of a de Sitter space.

\vspace{12pt}
\noindent

%\vfill

\newpage

\section{Towards quantum de Sitter geometry}
\label{sec:intro}

Quantum observables are needed in any candidate theory of nonperturbative quantum gravity to understand its 
dynamics on Planckian scales and explore its phenomenological consequences. Observables enable us to
describe the largely unknown properties of ``quantum geometry" in a Planckian regime in quantitative terms
and to verify the presence or otherwise of a classical limit. Alas, they are hard to come by. 
Already in classical general relativity, due to the requirement of diffeomorphism invariance, observables
tend to be highly nonlocal, e.g. given by the spacetime integral of some scalar quantity. 
Major difficulties in the construction of observables in nonperturbative quantum gravity theories are an appropriate 
implementation of diffeomorphism invariance and the fact that the continuum ingredients of the classical
theory, including a smooth metric field $g_{\mu\nu}(x)$, are usually not available. 
Instead, such formulations may be based on non-smooth or even discrete dynamical degrees of freedom.

The nonperturbative path integral approach to quantum gravity in terms of Causal Dynamical Triangulations (CDT)
lies at the less radical end of possible choices (see \cite{review1,review2} for reviews). 
At the regularized level, it works with ensembles of piecewise flat geometries, whose description does not require the 
introduction of coordinate systems and their associated redundancy. 
It is nevertheless close in spirit to classical gravity, in the sense that 
measurements of volumes and distances can be performed in the lattice setting in a straightforward way. However, 
in a continuum limit where the regulators are removed, 
these geometric notions can (and in explicit situations often do) scale non-canonically, that is, not according to the
``na\"ive" dimensionality expected from classical considerations. Such behaviour
can be a signature of genuine quantum gravity effects.  

Using the fact that CDT quantum gravity is amenable to Monte Carlo simulations, several quantum observables have been 
implemented and their behaviour investigated, with noteworthy outcomes. Perhaps the most spectacular
one is the emergence of an extended four-dimensional universe of de Sitter-type in one of the phases of the model  
\cite{CDT1,desitter}. Evidence for this comes from analysing the expectation value $\langle V_3(\tau)\rangle$
of the so-called volume profile, namely, the three-volume $V_3$ of the spatially compact universe as a function
of proper time $\tau$. This expectation value can be matched with great accuracy to the volume profile of a classical
de Sitter space\footnote{This is a {\it Euclidean} version of de Sitter space (the round four-sphere), since computations 
in CDT take place on a configuration space of Wick-rotated geometries, see \cite{review1} for details.}. 
Because the universes generated dynamically in the computer simulations turn out to be rather small (less than 20
Planck length across \cite{desitter,review1}), the quantum fluctuations $\delta V_3$ around the de Sitter profile are sufficiently large to 
be analyzed also. The behaviour of the lowest eigenstates extracted from the simulations 
matches well with the results of a semiclassical minisuperspace analysis in the
presence of a positive cosmological constant (which is the appropriate comparison for CDT). 
Remarkably, it is even possible to reconstruct the effective minisuperspace action for the three-volume (or equivalently,
the Friedmann scale factor) purely from the computer data, in the context of the so-called reduced CDT transfer matrix \cite{transfer}.
  
It should be emphasized that the continuum minisuperspace model one uses as a benchmark to establish
the presence of (semi-)classical behaviour of the quantum observable $\hat{V_3}$
is obtained by {\it assuming} homogeneity and isotropy of the universe at the outset and making a corresponding  
ansatz for the continuum metric, as is usual in cosmology. 
By contrast, the CDT set-up is background-independent and does {\it not} impose any symmetry assumptions 
on the geometries in the path integral. The fact that the particular observable $\hat{V_3}(\tau)$ agrees within
measuring accuracy with the corresponding quantity in a de Sitter universe is therefore a very nontrivial outcome.
Given the highly nonclassical nature of the quantum geometry one encounters 
in the limited range of scales currently accessible computationally,
it is fortunate that some kind of semiclassical behaviour has been observed at all. A plausible explanation for the
apparent robustness of the volume observable $\hat{V_3}$ is presumably its global nature. It does not imply 
that the same is true for other quantum observables. 

In view of the above one could be tempted to conclude that the quantum geo\-metry 
of CDT quantum gravity in the de Sitter phase\footnote{see \cite{review2} for an up-to-date 
description of the phase structure of CDT} $C_{dS}$
to first approximation simply {\it is} that of de Sitter space, perhaps in a suitably averaged or coarse-grained sense.
However, no such conclusion can be drawn from the behaviour of the scale factor alone. Even if there was evidence that
the quantum geometry could be thought of as approximating a smooth Riemannian manifold -- which presently there is not -- 
the scale factor is just one of the metric modes and its behaviour is
in principle compatible with many different kinds of local geometry.
On the other hand, it is conceivable that {\it if} some kind of semiclassical geometry emerges, approximate homogeneity
and isotropy will emerge alongside. The de Sitter phase is a natural region to look for such behaviour.
By contrast, the neighbouring bifurcation phase $C_b$, which also exhibits de Sitter-like volume profiles, is distinguished by
the appearance of a localised, string-like structure and therefore appears incompatible with homogeneity.  

If we had a well-defined measure of curvature in the quantum theory, it would allow us to 
probe the local geometry of the de Sitter universe and search for further evidence of semiclassical behaviour 
compatible with a constantly curved de Sitter space. Fortunately, a suitable notion of quantum curvature has become available
recently. This is not accidental, in fact, the new {\it quantum Ricci curvature} \cite{qrc1,qrc2} was developed 
with precisely this application in mind. Since the new observable by construction monitors the local geometry
at some prescribed length scale $\delta$, it will also be sensitive to the local quantum fluctuations at this scale,
which are not small. The crucial question is then whether we will be able to see {\it any} signal of semiclassical behaviour 
when measuring the expectation value of the quantum curvature. A possible outcome would be that the quantum fluctuations
in the near-Planckian regime under consideration are too large for this particular quantum observable to display any
recognizably classical features. A more exciting possibility would be to identify some (approximate) length threshold 
$\delta_0$ above which semiclassical behaviour is observed.

In assessing any results, one needs to take into account the usual limitations 
of the computational set-up with regard to the efficiency of the computer algorithm and the size of the simulated spaces
(in our case up to over a million building blocks).  
Like in lattice simulations of any four-dimensional quantum field theory, one faces the challenge of trying to extract reliable results from
performing measurements in a relatively narrow window of scales between short-distance lattice artefacts and finite-size effects
at large distances. As usual, this will be reflected in the presence of numerical and systematic uncertainties.

Previewing our main result, we have found evidence for the presence of semiclassical behaviour of the 
scalar quantum Ricci curvature that is compatible with a sphere of constant positive curvature. The signal appears
to be just about within the reach of our measurements, in the sense that it is only visible clearly when distances are
measured along the links of the dual lattice, for reasons that will be explained below. 
In view of the important role played by de Sitter space in standard cosmology, 
this further corroboration of the
de Sitter nature of the emergent quantum geometry in CDT quantum gravity presents a new and very promising 
result.\footnote{The relation between CDT quantum gravity and cosmology is discussed in \cite{cosmo,review2}.}

In the remainder of this article, we start in Sec.\ 2 by reviewing briefly the motivation and construction of the quantum Ricci 
curvature. Sec.\ 3 contains
a summary of the relevant ingredients of the CDT set-up. The curvature measurements and key results of our analysis
are described in Sec.\ 4. For the most part, they concern the expectation value of the quantum Ricci scalar, obtained by directional 
averaging, but we also present evidence that the full quantum Ricci curvature behaves the same way in space- and timelike directions. 
The final Sec.\ 5 contains a summary and outlook.

\section{Quantum Ricci curvature}
\label{sec:qrc}

The concept of quantum Ricci curvature arose from a quest for a notion of curvature applicable in
nonperturbative quantum gravity and the CDT approach in particular. Its starting point is the fact that on positively
curved Riemannian spaces two sufficiently close, small geodesic spheres are closer to each other than their respective centres, 
whereas the opposite is true for spaces of negative curvature. This classical observation underlies Ollivier's
construction of the ``coarse Ricci curvature" on metric spaces equipped with a random walk \cite{ollivier}. Ollivier-Ricci
cur\-vature was 
brought to our attention through a graph-theoretical implementation in models of emergent quantum spacetime 
in the work of Trugenberger and collaborators \cite{trugenberger}.

\begin{figure}[t]
\centerline{\scalebox{0.45}{\rotatebox{0}{\includegraphics{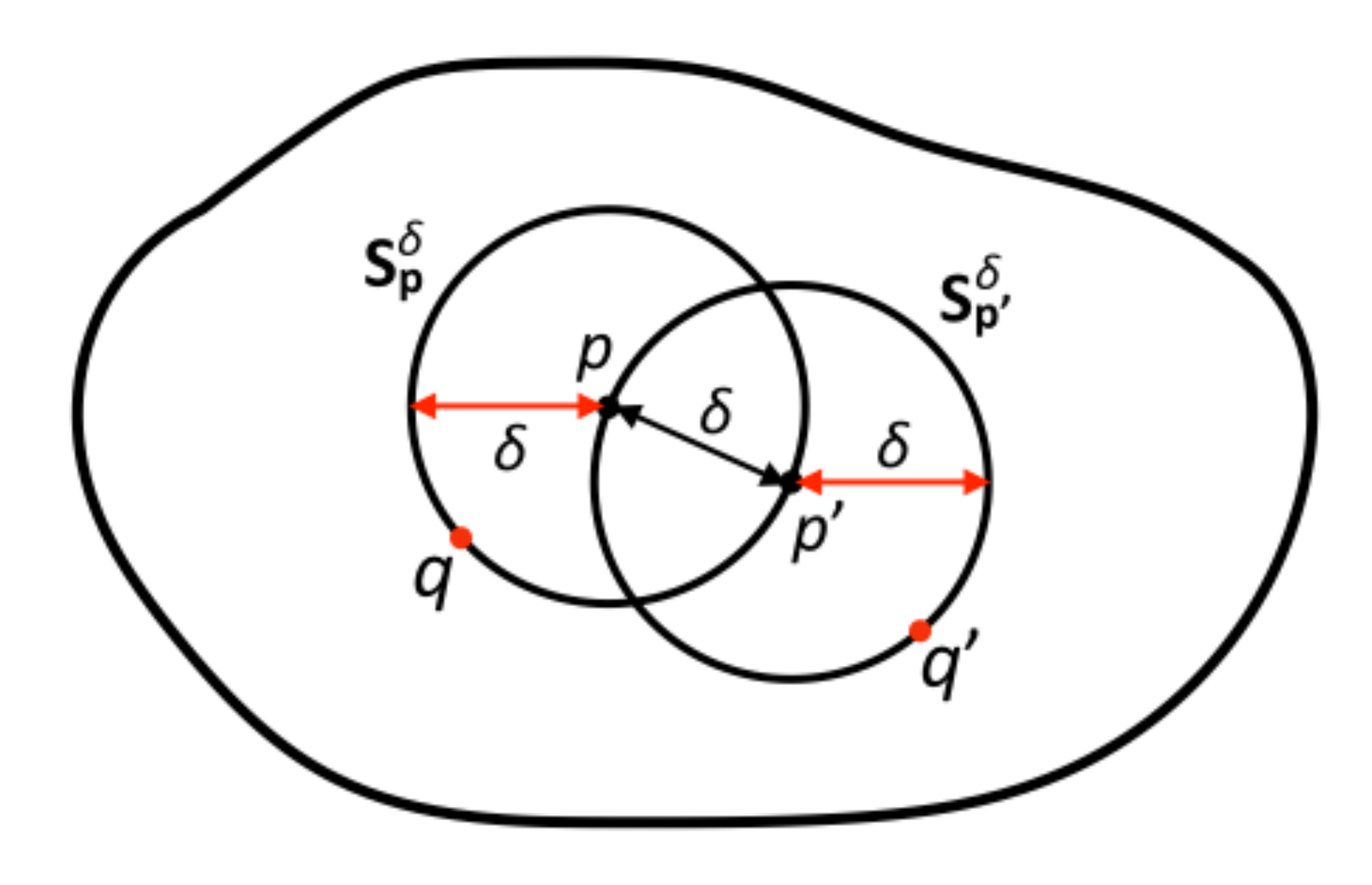}}}}
\caption{The local geometric set-up for determining the quantum Ricci curvature consists of two overlapping spheres
$S_p^\delta$ and $S_{p'}^\delta$ of radius $\delta$, whose centres $p$ and $p'$ likewise are a distance $\delta$ apart. 
The average sphere distance (\ref{sdist}) is obtained by averaging over the distances between all point
pairs $(q,q')$ along the two spheres.}
\label{fig:newintersect}
\end{figure}

Building on this idea, the {\it quantum Ricci curvature} $K_q$ introduced by us in \cite{qrc1} is an ingredient for constructing 
quantum observables that can be evaluated on ensembles of simplicial manifolds
like those appearing in Euclidean or Causal Dynamical Triangulations.\footnote{Calling it a ``quantum" curvature refers to the fact that
the primary motivation for its introduction was the nonperturbative quantum theory. However, it can just as well be implemented on classical
metric spaces \cite{qrc1,qrc2}.} Our choice of the quasi-local geometric 
set-up is that of two intersecting geodesic
spheres $S_p^\delta$ and $S_{p'}^\delta$ of radius $\delta$, with centres $p$ and $p'$ that are a distance $\delta$ apart,
as illustrated in Fig.\ \ref{fig:newintersect}.
The key quantity to be computed is the average sphere distance $\bar d$,
which we define as the double-sum or double-integral over all pairs of points $(x,x')\in
S_p^\delta \times S_{p'}^\delta$ of the distance $d(x,x')$.
Considering for illustration the case of a four-dimensional smooth Riemannian
manifold with metric $g_{\mu\nu}$, the average sphere distance $\bar{d}(S_p^{\delta},S_{p'}^{\delta})$ 
between the two three-spheres centred at $p$ and $p'$ is defined as the normalized double-integral
\begin{equation}
\bar{d}(S_p^{\delta},S_{p'}^{\delta}):=\frac{1}{vol(S_p^{\delta})}\frac{1}{vol(S_{p'}^{\delta})}
\int_{S_p^{\delta}}d^{3}x\; \sqrt{h} \int_{S_{p'}^{\delta}}d^{3}x'\; \sqrt{h'}\ d(x,x'),
\label{sdist}
\end{equation}   
where $h$ and $h'$ are the determinants of the metrics induced on $S_p^{\delta}$ and $S_{p'}^{\delta}$
(which are also used to compute the sphere volumes appearing in the formula), and geodesic distances are
calculated with respect to $g_{\mu\nu}$.
When implementing the average sphere distance on the triangulated manifolds of the CDT path integral, 
we use straightforward discrete analogues of distance and volume measurements, to be detailed below. 
Given a sphere distance $\bar{d}(S_p^{\delta},S_{p'}^{\delta})$, 
the quantum Ricci curvature $K_q(p,p')$ associated with the point pair $(p,p')$ 
is defined through the dimensionless quotient 
\begin{equation}
\frac{\bar{d}(S_p^{\delta},S_{p'}^{\delta})}{\delta}=c_q (1 - K_q(p,p')),\;\;\; \delta =d(p,p'),
\label{qric}
\end{equation}
where $c_q$ is a positive constant, which depends on the metric space under consideration,
and $K_q$ captures any nontrivial dependence on $\delta$. As explained in \cite{qrc1,qrc2},
$K_q(p,p')$ should be thought of as a generalization of the continuum Ricci curvature $Ric(v,v)=R_{ij}v^iv^j$,
the Ricci tensor evaluated on a vector $v$, with the vector substituted by a pair of spheres with centres 
$p$ and $p'$. In the continuum, the constant $c_q$ 
is given by $c_q=\lim_{\delta\rightarrow 0} \bar{d}/\delta$; on piecewise flat spaces, one has to define
a suitable analogue. The simplest way of converting the quantity (\ref{qric}) into
an observable is by integrating over all pairs $(p,p')$ of points that are separated by a distance $\delta$.
In this case, all directional information is integrated over, and one retains a quantity $K_q(\delta)$, which can 
be thought of as the average scalar curvature at length scale $\delta$ of the underlying metric space.  
On a curved classical manifold, this quantity will in general be impossible to compute, but in the quantum theory under 
consideration
the corresponding expectation value $\langle K_q(\delta)\rangle$ can be extracted readily by Monte Carlo simulation, 
subject to numerical uncertainties. Note that we will in general be interested in measuring this quantum curvature as a
function of the distance scale $\delta$, where $\delta$ is not necessarily small. As explained in the previous section,
one of the aspects we will be looking for is a possible onset of semiclassical behaviour on sufficiently large scales. 

After introducing the quantum Ricci curvature and thoroughly testing its behaviour on various piecewise flat
classical spaces in \cite{qrc1}, we presented the first quantum implementation in \cite{qrc2} for the case of
Euclidean (Liouville) quantum gravity in two dimensions, defined in terms of a nonperturbative path integral
of Dynamical Triangulations (DT). This well-studied model of low-dimensional quantum gravity
is an interesting testing ground for the quantum curvature because the nature of its quantum geometry is known to be fractal
and highly nonclassical \cite{dim2d,dim2dnum}. 
Somewhat remarkably, above short distances where lattice artefacts dominate 
we found that the quantum Ricci curvature can be matched with good accuracy to that of a five-dimensional continuum 
sphere of constant curvature. 
Since two-dimensional quantum gravity does not have a nontrivial classical limit, we do not have an immediate interpretation 
of this result,
or a way to assess whether it is physically sensible.\footnote{Given that the Hausdorff dimension of the model is 4, it might
have been less surprising to find the behaviour of a {\it four}-dimensional sphere. Since the
quantum Ricci curvature evaluated on round continuum spheres depends only weakly on the dimension,
we cannot exclude this possibility entirely. At any rate, extracting a dimensionality from sphere matching
does not appear very reliable, given numerical and systematic uncertainties, see also our discussion below.} However, 
it demonstrates a considerable degree of robustness of
the quantum Ricci curvature, in the sense of averaging out large local quantum fluctuations and producing a 
classical-looking outcome. 

In four dimensions, this averaging-out has not yet been observed for other notions of curvature one  
might consider in the nonperturbative quantum theory. One of them is the concept of local deficit angle \cite{regge},
which is well defined at the scale of the lattice cut-off, but has so far not led to interesting macroscopic
curvature observables in a continuum limit (see also the discussion in \cite{review2}, Sec.\ 7.2). Another one is the
gravitational Wilson loop of the Levi-Civita connection of the metric. At least for the large loops considered in
an investigation in four-dimensional CDT quantum gravity, its expectation value  
did not reveal any interesting relation with curvature \cite{wilsoncdt}. This does not preclude the
existence of such a relation for suitably averaged smaller loops (see also \cite{flux} for a related continuum discussion),
but this issue has not yet been analyzed in detail. By comparison, the quantum Ricci curvature has produced the most
promising results so far, providing a strong motivation for its investigation in fully-fledged four-dimensional
quantum gravity, which is the subject of the remainder of this article.

\section{CDT set-up in four dimensions}
\label{sec:ct4d}

Before turning to the measurements of the quantum Ricci curvature, we will summarize the main ingredients of the
CDT approach to the extent that is needed to understand the essence of our investigation. For detailed expositions of the
motivation and the technical and conceptional underpinnings of Causal Dynamical Triangulations 
we refer to the available reviews \cite{review1,review2}. 

The regularized gravitational path integral of CDT quantum gravity after Wick rotation takes the form
\begin{equation}
Z^{\rm CDT}_{eu}=\sum_{T} \frac{1}{C(T)}\ {\rm e}^{-S^{\rm CDT}_{eu}[T]},
\label{pintegral}
\end{equation} 
where the sum is taken over a set of triangulations $T$ with a well-defined causal structure, 
$C(T)$ is the order of the automorphism group of $T$,  
and the Euclideanized Einstein-Hilbert action $S^{\rm CDT}_{eu}$ is given by
\begin{equation}
S^{\rm CDT}_{eu}[T]\!=\! - \kappa_0 N_0(T) + \Delta (2 N_{41}(T)+N_{32}(T)-6 N_0(T))+\kappa_4 (N_{41}(T)+N_{32}(T)).
\label{action}
\end{equation}
The simple functional form of the gravitational action -- depending only on ``counting variables"\footnote{$N_0$
counts the number of vertices (``$0$-simplices") in $T$, $N_{41}$ the number of four-simplices of type $(4,1)$ and
$N_{32}$ the number of four-simplices of type $(3,2)$; adding them one obtains the total number of four-simplices,
$N_4:=N_{41}+N_{32}$.} -- comes from the fact 
that the simplicial manifolds $T$ are
assembled from just two types of four-dimensional building blocks, namely, four-simplices of type $(4,1)$ and of
type $(3,2)$. These two types come from two different flat Minkowskian building blocks before the analytic continuation to
Euclidean signature, and have different numbers of time- and spacelike edges. The bare coupling constants
appearing in the action (\ref{action}) are the inverse Newton constant $\kappa_0$, the cosmological constant $\kappa_4$
and the asymmetry parameter $\Delta$, which captures the finite relative scaling between the geodesic lengths of
time- and spacelike edges and becomes a relevant coupling in the nonperturbative regime. 

Recall that CDT configurations come with a discrete notion of proper time, which is used to enforce a version
of global hyperbolicity on the simplicial mani\-folds. A triangulated manifold $T$ can be thought of as a discrete sequence of 
three-dimensional equilateral triangulations, each representing the geometry at a fixed (integer) moment in proper time. 
Four-dimensional ``layers" of spacetime, made from $(4,1)$- and $(3,2)$-simplices, interpolate between 
neighbouring spatial slices with time labels $t$ and $t\! +\! 1$. 
The topology of the spatial slices is fixed, usually to that of a three-sphere, which will also be our choice here. 
Because of the ``$(3+\! 1)$-dimensional'' character of a given simplicial manifold in the CDT ensemble, 
all of its vertices lie in one of the spatial slices and therefore
have an integer time label $t$. 

For the convenience of the simulations, we work with a compactified time
direction, leading to an overall topology of $S^1\!\times\! S^3$. By making the time direction long (with $t_{tot}\! =\! 120$ time
steps), this choice should have little influence on the measurement results of a quasi-local quantity like the
quantum Ricci curvature.
However, like for all choices of global boundary conditions,
one needs to pay close attention to the possibility of associated finite-size effects in any given numerical experiment.
As usual in CDT, we will take data at different fixed spacetime volumes, measured in terms of the counting
variable $N_{41}$, and then use finite-size scaling to extrapolate to infinite discrete volume. 
Our measurements have been conducted in the range $N_{41}\!\in\! [75k, 600k]$, corresponding to an
approximate range $N_4\! \in\! [150k,1.200k ]$. 
To conduct measurements at a desired target volume $\bar{N}_{41}$, we fine-tune the value of the cosmological constant 
such that the average volume coincides with $\bar{N}_{41}$. To make sure that the simulations stay in a narrow window
around this value, we add a volume-fixing term $\epsilon(N_{41}(T)\! -\! \bar{N}_{41})^2$ to the bare action, 
with $\epsilon=0.00002$. 
Note that we only take into account measurements taken at the exact value $\bar{N}_{41}$. 
We have fixed the two remaining free constants in the action to $(\kappa_0,\Delta)=(2.2,0.6)$, 
corresponding to a point in the phase diagram that has been used in previous investigations of CDT quantum gravity.
This puts us inside the so-called de Sitter phase $C_{dS}$ already mentioned in Sec.\ \ref{sec:intro}.
In the simulations, we have proceeded by Monte Carlo sampling, using an adapted version of a code originally written by A.\ G\"orlich.
We used $7.5\times 10^7$ suggested Monte Carlo updates per sweep, performing one measurement at the end of each sweep, 
and an initial $80k$ of sweeps for thermalization. 
To speed up the thermalization process we used thermalized configurations generated at smaller sizes as initial seeds for 
the larger configurations.

\begin{figure}[t]
\centerline{\scalebox{0.65}{\rotatebox{0}{\includegraphics{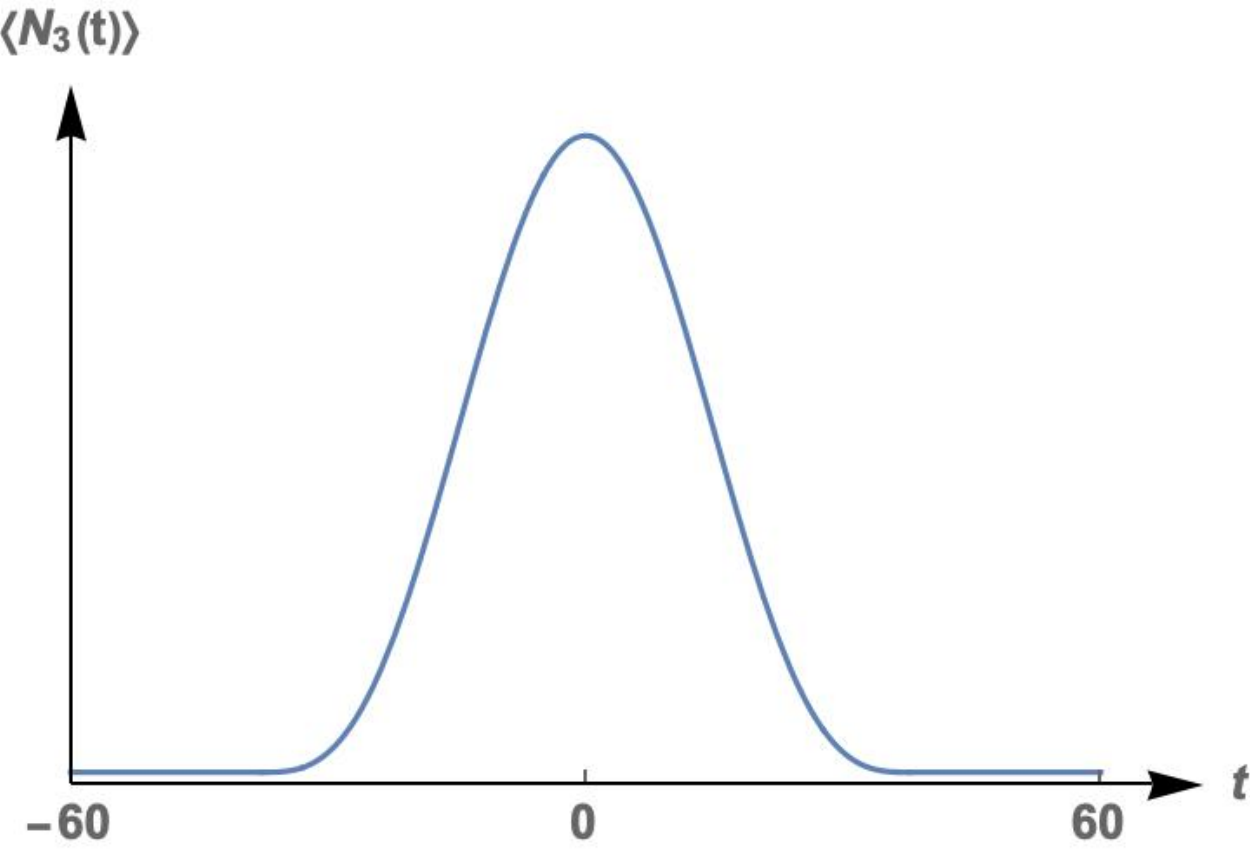}}}}
\caption{Schematic illustration of the volume profile $\langle N_3(t) \rangle$ of the quantum spacetime
in the de Sitter phase, consisting of a spatially extended universe (here centred around $t=0$) and a stalk of minimal spatial extension.}
\label{fig:desitterprofile}
\end{figure}
Recall from the introduction that one of the hallmarks of the de Sitter phase is the presence of a volume profile 
matching that of a Euclidean de Sitter universe. Its typical shape is illustrated schematically by Fig.\ \ref{fig:desitterprofile}: 
in a connected interval along the time axis, the expectation value follows that of a
de Sitter universe with volume profile $\langle N_3(t) \rangle\propto \cos^3 (t/const)$,
where $N_3(t)$ (counting the number of three-simplices at integer $t$) is the discrete counterpart of the
continuum volume $V_3(\tau)$. The remainder of the time axis is taken up by a thin ``stalk", whose slices at integer $t$
have the minimal spatial volume compatible with a simplicial manifold of $S^3$-topology.\footnote{Since the
stalk width is of cut-off size, its volume in any continuum limit will vanish. This points to a remarkable example of dynamical emergence:
within a configuration space of geometries of topology $S^3\!\times\! S^1$, the quantum dynamics drives the
system to a state where its effective topology is that of a four-sphere (see \cite{cosmo} for a related discussion).}    
In our simulations, a stalk is always present, even for maximal volume $N_{41}$. In other words, the time extension is 
sufficiently large to fit the de Sitter universe in its entirety. For simplicity, we treat the stalk region in the same
way as the bulk of the geometry in the measurements. This is justified on the grounds that its (spacetime) volume
is a very small fraction of that of the bulk, which furthermore decreases with increasing values of the volume. 
If we define any spatial slice with volume $N_3\!\leq\! 100$ as belonging to the stalk\footnote{Typical spatial volumes in the stalk lie
in the interval $N_3\!\in\! [5,30]$.}, the stalk makes up for 
about 4\% of the total volume at $N_{41}\! =\! 75k$ and 0.35\% at $N_{41}\! =\! 600k$. At the same time,
the universe -- defined as the complement to the stalk -- occupies roughly 24 of the 120 time steps at $N_{41}\! =\! 75k$
and about 47 of the 120 time steps at $N_{41}\! =\! 600k$.
We have also made sure that when measuring sphere distances, a crucial ingredient in the 
curvature analysis, the periodic identification of time does not give rise to the presence of unintended shortcuts running
through the stalk.

\section{Measuring the quantum Ricci curvature}
\label{sec:meas}

Besides specifying how precisely the quantum Ricci curvature in the four-dimen\-sional CDT path integral is measured,
an important question is what the result is going to be compared to. The classical reference spaces for which the quantum Ricci
curvature as a function of scale has been computed exactly are maximally isometric Riemannian spaces of constant curvature,
i.e.\ spheres, hyperbolic spaces and flat space. In each of these three cases, the quotient $\bar{d}/\delta$ of eq.\ (\ref{qric}) displays a 
distinctly different behaviour \cite{qrc1}.
In terms of genuine nonperturbative quantum systems, the only reference point at this stage is the investigation of two-dimensional
Euclidean quantum gravity already mentioned in Sec.\ \ref{sec:qrc} \cite{qrc2}. Four-dimensional quantum gravity is expected to be
very different from
this toy model. It should involve at least two distinct physical scales, a microscopic one associated with quantum fluctuations in a Planckian
regime and a macroscopic one related to the size of the universe. In addition, it should have a well-defined classical limit
containing local, gravitonic field excitations. As explained earlier, we are looking for further corroborating evidence for the existence 
of this classical limit and in particular for the de Sitter nature of the dynamically generated quantum geometry in the CDT formulation. 
 
In what follows, we will work with two well-known discrete notions of geodesic distance, the link distance and the dual link
distance. The choice of one or the other amounts merely to choosing a particular discretization, which is part of the regularized 
set-up and should not affect any continuum results.
However, due to the numerical limitations of the simulations, it turns out that for the particular
observable under consideration one of them is clearly preferable, for reasons that will become apparent. 

The link distance depends on two vertices $x$ and $x'$ of the triangulation $T$ and is defined as the length of the shortest
path connecting $x$ and $x'$ along the links of $T$, in discrete units (i.e. counting the number of links). 
The dual link distance is defined analogously between
pairs of {\it dual }vertices $x$ and $x'$, which by definition are the centres of four-simplices of $T$. 
It is given by the discrete length of the
shortest path between $x$ and $x'$ consisting of dual links, where a dual link is a straight line segment connecting the centres of two 
adjacent four-simplices. Using the same notation $d(x,x')$ for either of these integer-valued geodesic distances, 
the average sphere distance on piecewise flat CDT geometries takes the form  
\begin{equation}
\bar{d}(S_p^{\delta},S_{p'}^{\delta})=\frac{1}{N_0(S_p^{\delta})}\frac{1}{N_0(S_{p'}^{\delta})}
\sum_{x\in S_p^{\delta}} \sum_{x'\in S_{p'}^{\delta}} d(x,x'),\;\;\;\; d(p,p')=\delta,
\label{sdist_d}
\end{equation}   
where the ``$\delta$-spheres" $S_p^{\delta}$ and $S_{p'}^{\delta}$ consist of all (dual) vertices at (dual) link distance $\delta$
from the (dual) centre vertex $p$ and $p'$ respectively, and $N_0$ counts the number of (dual) vertices. 
One can think of $\bar{d}(S_p^{\delta},S_{p'}^{\delta})$ as a generalization of a two-point function, where the two ``points" have been
substituted by local spheri\-cal regions.
Unlike what happens on a constantly curved Riemannian space, say,
a $\delta$-sphere does not in general define a (simplicial version of a) three-sphere, topologically speaking, but will typically
consist of several such components, which moreover can ``touch" each other in complicated ways. This well-known feature
appears even for small $\delta$ (beyond $\delta\! =\! 1$), and has to do with the highly irregular nature of generic quantum
configurations. It is an entirely expected property of the nonperturbative quantum theory and in no way contradicts
the emergence of semiclassical behaviour once we study expectation values of observables on sufficiently large scales. 

We measured the expectation value of the average sphere distance using a Metropolis algorithm. 
Our set of ergodic moves consists of slightly altered Pachner moves conserving the sliced structure.\footnote{The moves combine 
the Pachner moves in a nonstandard fashion to reduce correlation times.} By demanding that the probability for 
accepting a move obey the condition of detailed balance based on the action (\ref{action}), we 
make sure that after thermalization the probability of a given configuration will correspond to its probability in the state sum. 
To determine the expectation value of the average sphere distance then only requires a simple averaging over measurements.
For fixed volume $N_{41}$, we performed a double-sampling of the ``two-point function"    
$\bar{d}(S_p^{\delta},S_{p'}^{\delta})$ of eq.\ (\ref{sdist_d}) over point pairs $(p,p')$ and geometries $T$.
From this, we extracted the (normalized) expectation values $\langle \bar{d} (\delta)\rangle/\delta$ of the average sphere 
distance as a function of $\delta$. 
A single measurement proceeded in
the following steps: (i) pick a random vertex $p\!\in\! T$ in the triangulation $T$ that has been generated at this stage of
the Metropolis algorithm; (ii) pick a vertex $p'$ randomly from the
sphere of radius $\delta\! =\! 1$ around $p$ and compute $\bar{d}(S_p^{\delta},S_{p'}^{\delta})/\delta$; (iii) repeat step (ii) 
for the same $p$ and a randomly chosen point $p'$ from the $\delta$-sphere around $p$, for $\delta\! =\! 2,3,\dots,15$, yielding
a total of 15 data points for $T$. 
After thermalization we performed 20.000 measurements for each system size. The measurements slowed down considerably for the 
larger system sizes, which meant that for $N_{41}\! =\! 600k$ we had to run simulations for a total of three months to reach the desired 
number of measurements.

\begin{figure}[t]
\centerline{\scalebox{0.75}{\rotatebox{0}{\includegraphics{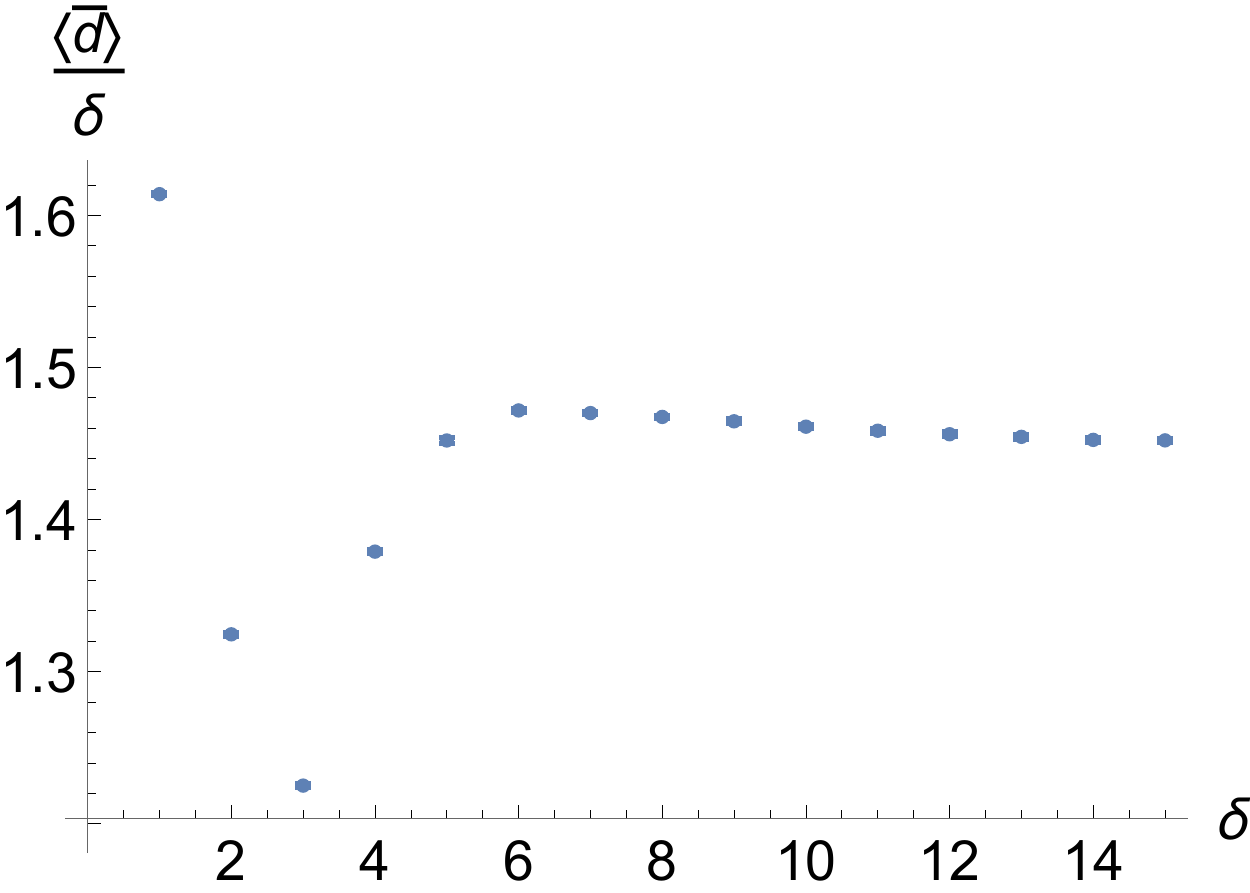}}}}
\caption{Normalized average sphere distance $\langle\bar{d}\rangle/{\delta}$ 
as a function of $\delta$, measured at volume $N_{41}\! =\! 480k$ and using the link distance. (Error bars too small to be shown.)
}
\label{fig:linkd}
\end{figure}

We begin with a discussion of the curvature measurements using the link distance. 
We have measured the expectation value $\langle \bar{d}\rangle/{\delta}$ of the normalized 
average sphere distance at various fixed volumes. The results for the largest of these volumes, $N_{41}\! =\! 480k$, and
for $\delta\in [1,15]$ are shown
in Fig.\ \ref{fig:linkd}; the situation for smaller volumes is qualitatively similar. 
On the basis of previous investigations \cite{qrc1,qrc2}, one expects to find lattice artefacts in the region of small $\delta \lesssim5$. 
We observe the familiar short-distance overshooting for the smallest $\delta$-values, followed by a dip at $\delta\! =\! 3$ (a feature
that has not been seen before), before
the curve settles to an approximately constant value $\approx\! 1.46$. Since the non-constant behaviour is confined to the 
region dominated by short-distance discretization effects, this appears to indicate a universe of vanishing curvature,
contrary to our expectation of finding a de Sitter behaviour with positive curvature.  
Even when taking this outcome at face value, a slightly puzzling aspect is the small value 
of the constant $\langle \bar{d}\rangle/{\delta}$, which
is indicative of a low-dimensional ($\leq\! 2$) rather than a higher-dimensional space, judging by the results on regular lattices and
Delaunay triangulations \cite{qrc1} and in two-dimensional CDT quantum gravity \cite{brulo}.

To understand the origin of this behaviour better, we examined the nature of the $\delta$-sphere $S_p^\delta$ around a given 
vertex $p$, which is an important ingredient in the sphere distance measurements. It turns out that its shape is rather anisotropic 
in time and space directions when using the link distance to define it. Let us look at a typical sequence of three-spheres $S_p^\delta$,
$\delta\! =\! 1,2,3,\dots$, around a generic initial point $p$, where $p$ lies in a spatial slice with discrete time label $t_0$, and 
sufficiently far away from the stalk. 
One finds that after a few steps in $\delta$, the vast majority of the vertices in a given $\delta$-sphere lie either in the spatial slice at time 
$t_0+\delta$ or
in the spatial slice at time $t_0 -\delta$. While this is not necessarily problematic in itself, it does stand in the way of performing the
sphere distance measurements as intended. Because of the finite size of the triangulations we consider, 
the vertices of the sphere $S_p^\delta$ that lie in a spatial slice with time label $t_0\pm\delta$ will not just outnumber vertices
with time labels closer to $t_0$, but will already for rather small $\delta$ fill the entire slice at $t_0\pm\delta$. 
Once this happens at a given value $\delta$, the sphere of radius $\delta +1$ will in turn saturate the slices at times $t_0\pm (\delta +1)$.
This implies that the spheres will seize to grow as a function of $\delta$, and the distance between
pairs of points on two $\delta$-spheres will typically be given by the difference between their time labels. 
Following this reasoning, one would expect that above a threshold value for $\delta$ the sphere distances will behave effectively
like on a one-dimensional space, without any curvature.\footnote{On a one-dimensional continuum space,
the normalized sphere distance is a constant, $\bar{d}/\delta\!  =\! 1.5$.} This would provide a qualitative explanation for
the measurements of the expectation value $\langle\bar{d}\rangle/{\delta}$ shown in Fig.\ \ref{fig:linkd}, at least for large
values of $\delta$. 

This behaviour can be related to the distribution of vertex orders on the CDT ensemble of configurations, where the order of a vertex $p$
is defined as the number of edges meeting at $p$. 
Although the average vertex order is in the low twenties and its distribution is monotonically decreasing, 
vertices of high order (several hundreds) occur regularly, even at small volume.
Every time such a vertex is encountered by a $\delta$-sphere, it will in the next step grow to all 
of its connected neighbours, thereby contributing to the saturation of spatial slices described in the previous paragraph. 
By random sampling a few configurations, we found that a typical value at which this saturation happens is
$\delta\!\approx\! 7\pm 1$ at volume $N_{41}\! =\! 600k$, and $\delta\!\approx\! 6\pm 1$ at volume $N_{41}\! =\! 480k$, supporting our
argument vis-\`a-vis the data presented in Fig.\ \ref{fig:linkd}.

The fact that $\delta$-spheres start wrapping around entire spatial slices, leading to an effective one-dimensional behaviour, is an
undesirable finite-size effect, which implies that we are probing global topological properties rather than quasi-local geometry. 
Regarding our attempt to perform curvature measurements, it means that even for the largest volume considered, $N_{41}\! =\! 480k$, 
there is no appreciable range in $\delta$ between the region of discretization artefacts and the onset of finite-size effects where we could
perform any meaningful fits to the data. It does not seem possible to circumvent this problem with our currently available computer resources. 

Fortunately, there is a way out that has enabled us to make progress nevertheless. Key to alleviating the issues just discussed is using
the dual link distance instead of the link distance when computing the average sphere distance (\ref{sdist_d}). Note that the vertex order of a
dual vertex of a four-dimensional triangulation is always five. As a result, there are no high-order vertices that can act as 
``super\-spreaders" and lead to big jumps in the sphere volume from one $\delta$-sphere to the next. At the same time, the five dual links
emanating from a given dual vertex are maximally distributed over directions, unlike the time- and spacelike links of
the triangulation itself, which play rather different roles. In particular, to move from a given dual vertex $p$ in a spatial slice of constant
time\footnote{This requires an assignment of non-integer time labels to the dual vertices, which lie in between spatial slices of
integer $t$. We will define one such assignment later in this section.} $t$ to 
a dual vertex in an adjacent slice at time $t\!\pm\! 1$ requires at least four steps (moving along dual links) on the dual lattice, and there
may even be no such vertex in a neighbourhood of radius 4. By contrast, starting from a vertex $p$ of the original triangulation, there
are always timelike links connecting it to vertices with time labels $t\!\pm\! 1$. Since on the dual lattice the progression in timelike directions
is much more in line with the progression in spacelike directions, one may expect that the finite-size distortions are pushed out to 
larger values of $\delta$, making the curvature measurements more accessible. 

\begin{figure}[t]
\begin{tabular}{ll}
\includegraphics[width=0.48\textwidth]{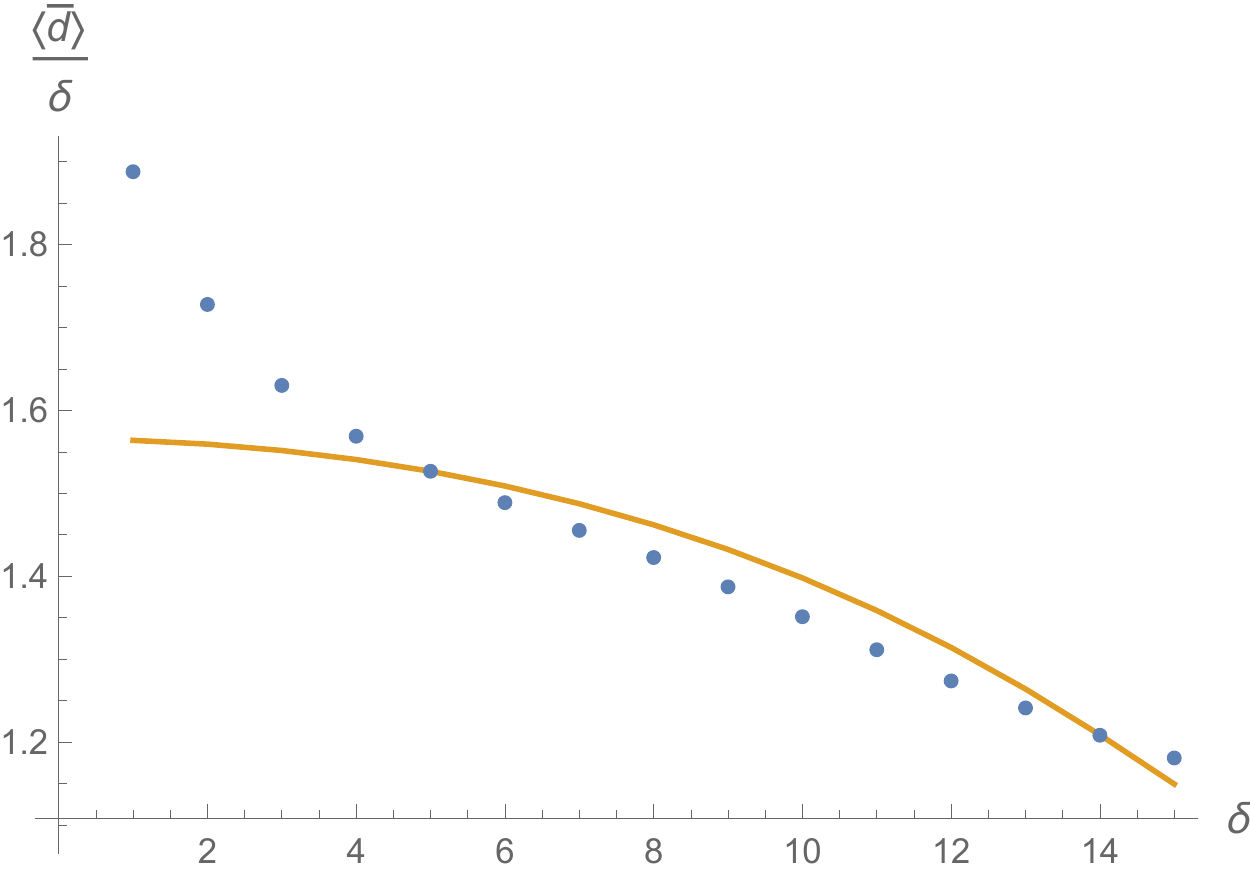}
&
\includegraphics[width=0.48\textwidth]{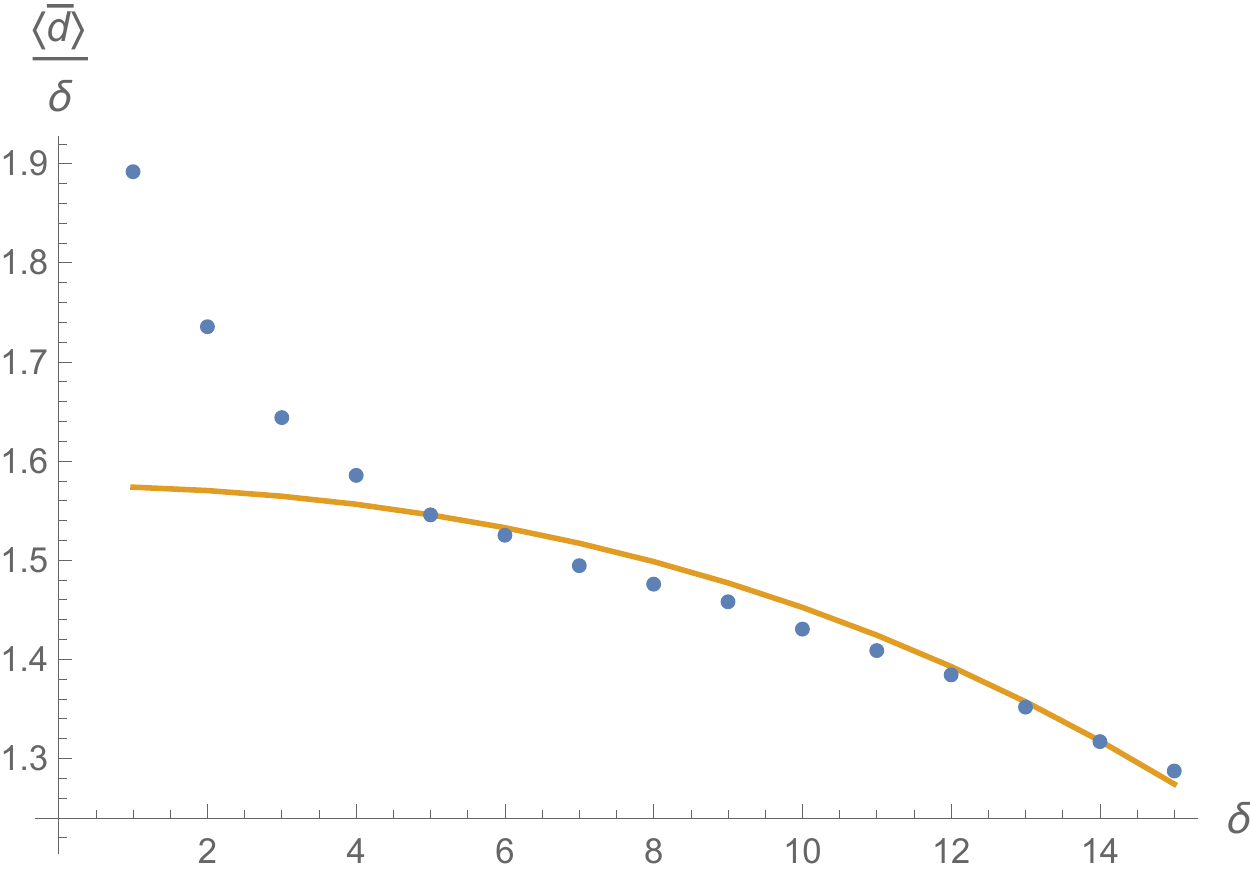}
\end{tabular}
\caption{Normalized average sphere distance $\langle\bar{d}\rangle/{\delta}$ 
as a function of $\delta$, measured at volume $N_{41}\! =\! 75k$ (left) and $N_{41}\! =\! 600k$ (right), using the dual link distance.
Best fits to a four-dimensional continuum sphere with radius $\rho\! =\! 12.09$ and $\rho\! =\! 13.96$ respectively are included 
for comparison. (Error bars are smaller than dot size.)
}
\label{fig:duallinkd}
\end{figure}

This expectation is borne out by our measurements of the normalized average sphere distance on the dual lattice. 
Fig.\ \ref{fig:duallinkd} shows the expectation values $\langle\bar{d}\rangle/{\delta}$ 
as a function of the dual link distance for the CDT ensembles of the smallest
and the largest volume we have considered, $N_{41}\! =\! 75k$ and $N_{41}\! =\! 600k$ respectively. 
There is now a clear, nontrivial 
dependence on $\delta$ beyond the initial discretization region, with the decrease indicating positive curvature! A second feature that
is rather striking is the resemblance with earlier measurements in two-dimensional (Euclidean) Liouville quantum gravity in terms of
Dynamical Triangulations \cite{qrc2}. 
As reported in Sec.\ \ref{sec:qrc} above, the data in that case could be matched to those of a five-dimensional continuum
sphere. 

More specifically, for both four-dimensional CDT and two-dimensional DT, 
after an initial region $\delta\lesssim 5$ with a rather steep overshoot, which we discard because of discretization artefacts, 
the data points 
start following a curve with a gentler decline. For small volumes, this decline is almost linear, as illustrated by the graph on the left in 
Fig.\ \ref{fig:duallinkd}, and not fitted well by the curve of a continuum sphere.\footnote{Note that fitting
to the continuum curves for the spheres involves a vertical shift, which is needed because of the non-universal
character of the constant $c_q$ in eq.\ (\ref{qric}). We have used an additive shift in Fig.\ \ref{fig:duallinkd}; a multiplicative shift 
leads to similar results, with slightly lower values of the radius $\rho$ \cite{thesis}.}
However, for increasing total volume $N_{41}$
the curve part $\delta\!\in\! [5,15]$
becomes gradually more convex, characteristic of the behaviour of a positively curved continuum sphere of radius $\rho$. 
This trend is clearly visible in the CDT data and mirrors the behaviour of two-dimensional DT with increasing volume $N_2$
(the number of triangles), but the approach to spherical behaviour is slower than in the DT case. 
Likewise, the quality of the sphere fit at the largest volume $N_{41}\! =\! 600k$ is not as good as the corresponding fit at the 
largest volume $N_2\! =\! 240k$ for the DT measurements. This is not particularly surprising since
on general grounds one would expect that a four-dimensional system shows a slower convergence as
a function of the total volume than a two-dimensional one, even if by some measures two-dimensional quantum gravity 
has an effective dimensionality larger than two.\footnote{Its Hausdorff dimension is 4, while its 
spectral dimension is 2 \cite{dim2d,dim2dnum,ambou}.}

\begin{figure}[t]
\centerline{\scalebox{0.75}{\rotatebox{0}{\includegraphics{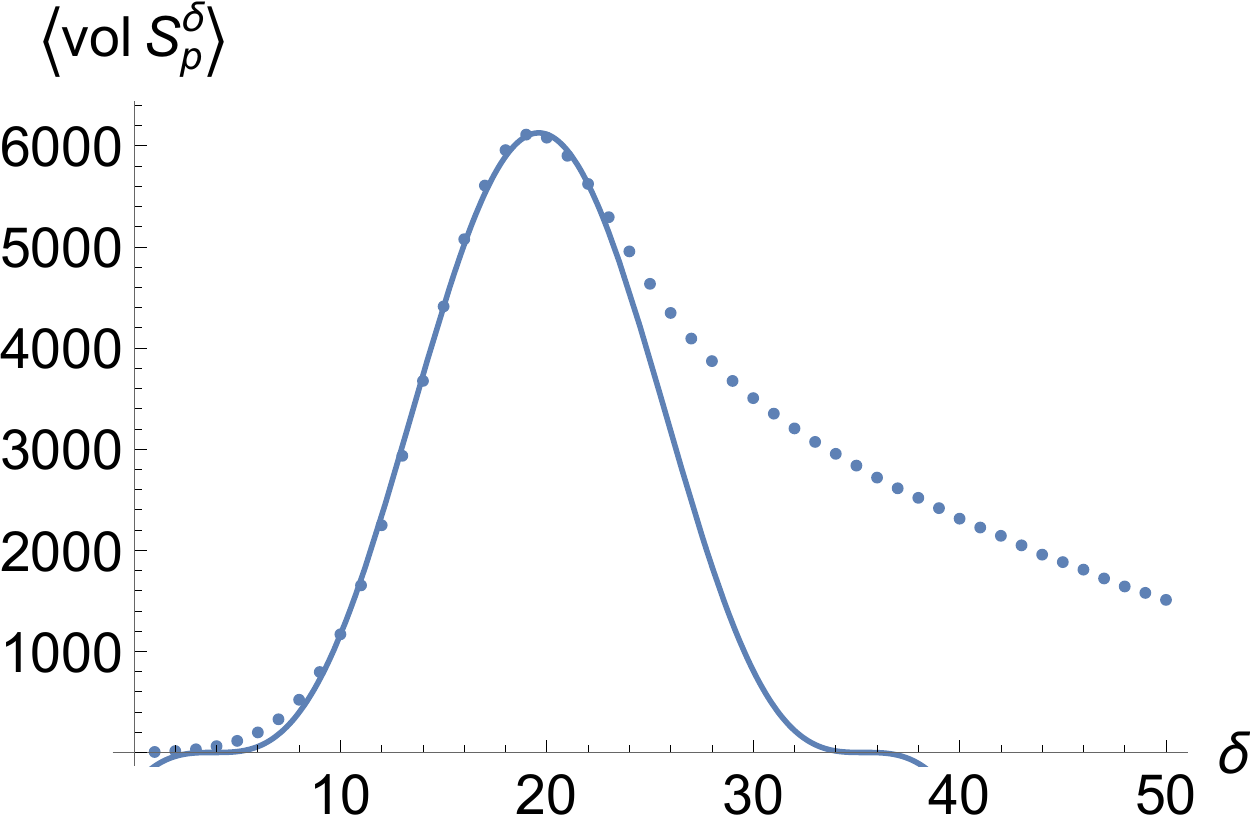}}}}
\caption{The average shell volume $\langle vol(S_p^\delta)\rangle$ as a function of the dual link distance $\delta$, at
volume $N_{41}\! =\! 75k$. The $\delta$-range where the data points are qualitatively well approximated by a $\sin^3$-function 
(continuous curve) gives us a rough estimate of the size of the region where the shell volumes display four-sphere behaviour.
Plots at higher volume look similar. The fit shown is to $6127 \sin^3 (0.10 \delta\! -\! 0.38)$ (including a horizontal offset),
corresponding to an effective radius $\rho_{\it eff}\! =\! 10$. 
}
\label{fig:profile}
\end{figure}

An additional criterion for estimating the $\delta$-range where the collected data may be considered reliable comes from measuring the volumes 
$vol(S_p^\delta)$ of $\delta$-spheres around an arbitrary dual vertex $p$
as a function of $\delta$. For sufficiently small $\delta$, their expectation values can be fitted well to a
$\sin^3 (\delta/\rho)$-curve, which would be the expected behaviour if the quantum universe was a round four-sphere of radius 
$\rho$ on all scales (see Fig.\ \ref{fig:profile} for illustration).  
However, as was observed previously \cite{geometry,Gorlichthesis}, the actual sphere volume plots 
display ``fat tails" for large $\delta$, leading to an asymmetry around the peak of the curve and a systematic deviation from a
$\sin^3$-behaviour for large $\delta$, not all of which appears to be attributable to finite-size effects.\footnote{We thank A.\ G\"orlich and
J.\ Gizbert-Studnicki for a discussion of this point.}
Since our present interest is in the quasi-local curvature properties of the quantum universe, 
we will not perform a more detailed analysis of the origin of this large-distance behaviour. 
Instead, the fit illustrated by Fig.\ \ref{fig:profile} allows us to make a rough estimate of the distance scale up to which 
no significant deviations from a pure four-sphere behaviour are observed, namely, up to $\delta$-values slightly beyond the
peak of the curve. 
Taking into account that the sphere distance measurements involve a configuration of two overlapping three-spheres with a combined linear 
extension of $3\delta$, an estimate for a lower bound of the region where the effects of large-scale non-sphericity {\it may} 
start manifesting themselves is $\delta\! \approx\! 8$ for $N_{41}\! =\! 75k$ and
$\delta\!\approx\! 13$ for $N_{41}\! =\! 600k$. This provides additional reassurance that -- unlike for the measurements we
performed using the link distance -- there is now a nontrivial range of $\delta$-values where we can meaningfully compare the curvature data to
those of continuum spaces, at least for the systems with sufficiently large volumes $N_{41}$. Lastly, note that our rather crude way
of extracting an effective four-sphere radius $\rho_{\it eff}$ from fitting shell volumes at $N_{41}\! =\! 75k$ (Fig.\ \ref{fig:profile}) 
produces a result in reasonable agreement with the radius $\rho$ extracted from the Ricci curvature measurements (Fig.\ \ref{fig:duallinkd}, left),
a coincidence that is also observed at larger volumes \cite{thesis}. This provides another consistency check that the curvature properties we
observe are indeed those of a four-sphere. 

On the basis of the numerical data collected, we conclude that there is strong evidence that the quantum geometry associated with
the de Sitter phase in CDT quantum gravity has positive average quantum Ricci curvature, as defined through the normalized average 
sphere distance (\ref{qric}).  
More specifically and quite remarkably, despite the fact that we are probing curvature properties at physical scales of no 
more than about 10 Planck
lengths \cite{desitter,review1}\footnote{Note that the discrete lattice distances we use here differ from those in references 
\cite{desitter,review1}, which were extracted from the lattice four-volume and the ordinary link distance. In general, one would expect that 
using the dual link distance leads to larger discrete distances for the same physical distance.
This also appears to be the case here.}, the expectation value of the Ricci scalar appears to be compatible with that of a four-sphere. 
It is tempting to conjecture that the discrepancy between the measured data and the corresponding
continuum curves (Fig.\ \ref{fig:duallinkd}) is a finite-size effect that will disappear for larger system sizes, similar to what
happens for DT in two dimensions. Another interesting possibility would be that part of the discrepancy  
is due to genuine quantum effects at short distances. In that case, a match with the classical curve may only occur for larger values of
$\delta$. We have not been able to explore any of these possible scenarios further, because the computational resources needed to go to system 
sizes much larger than $N_{41}\! =\! 600k$ are substantial and not currently at our disposal. 

Our motivation for comparing the sphere distance measurements to those on a {\it four}-dimensional sphere is that in CDT,
unlike in the DT toy example, we already have independent evidence that the dynamically generated quantum universe 
resembles a specific solution to the Einstein equations in four dimensions.
The quality of the CDT data does not allow us to make an independent
determination of the dimension of the sphere purely from the curvature measurements: on the one hand, 
the function $\bar{d}/\delta$ depends only weakly on the dimension of the underlying continuum sphere \cite{qrc2},
and on the other hand, it is apparent that even at the largest volume $N_{41}$ the data still display a systematic deviation 
from the sphere curves.\footnote{A more detailed
analysis of this issue, using and extending the finite-size scaling methods of \cite{qrc2}, can be found in \cite{thesis}. It is also
shown there that using a fitting range different from $\delta\! \in\! [5,15]$, to reduce possible short-distance or finite-size effects, 
does not sharpen or significantly alter the conclusions of the present work. Removing the data points of the
largest $\delta$-values leads to lower values for the effective radius $\rho$.}

Lastly, let us present an exploratory study which makes use of the fact that the quantum Ricci curvature has an inherent dependence on direction, 
given in terms of the relative position of the two spheres whose distance is measured. The easiest property associated with 
the local ``direction" on a
CDT configuration is its time- or spacelike character. This refers to a property inherited from the original, unique Lorentzian
piecewise flat spacetime before it was analytically continued to the corresponding configuration with Euclidean signature (see \cite{review1,review2} 
for more information on the Wick rotation in CDT). To keep things simple, we will examine the average sphere distance for two
extreme cases, where the distance between the two sphere centres $p$ and $p'$ is either maximally spacelike or maximally timelike, 
in the sense to be defined below.

Recall that we are working on the dual lattice, whose vertices do not carry a natural integer time label. For our present purpose, we will assign
certain fractional time labels\footnote{Similar fractional time labels were used in \cite{semi2} to measure ``refined" volume profiles.} to 
these vertices, which are motivated by the natural sequence of simplicial
building blocks one has to pass through to move forward in time along dual links. Starting at the centre of a four-simplex of type
(4,1), say, the shortest possible path along dual links to a (4,1)-simplex in the next time slice has length 4, and runs through a sequence
of building blocks given by $(4,1) \to (3,2) \to (2,3) \to (1,4) \to (4,1)$.\footnote{For the sake of this particular argument, a four-simplex of 
type $(m,n)$ is one that shares $m$ simplices with the spatial slice at some integer time $t$ and $n$ simplices with the spatial slice at
time $t\! +\! 1$. Note that elsewhere we do not distinguish between time orientations; for example,
$N_{41}$ refers to the combined number of simplices of type (4,1) and (1,4).} Correspondingly, we define ``dual time" to come in steps of
$1/4$. Assigning dual time $t_0\! =\! 0$ to the first (4,1)-building block in the sequence just mentioned, the corresponding fractional time labels
would read $0\to 1/4\to 1/2\to 3/4\to 1$, i.e. $t\! =\! t_0+n/4$ until the next (4,1)-simplex is reached.
Note that for any given $(4,1)$-simplex at time $t_0$ in a given triangulation, there is no guarantee that a shortest path of length 4 
to some $(4,1)$-simplex at time $t_0\!\pm\! 1$ exists. 

We will say that two sphere centres $p$ and $p'$ at dual link distance $\delta$ 
are ``maximally spacelike separated" if their fractional time labels are equal and ``maximally timelike separated" if their fractional
time labels differ by $\delta/4$. We performed sphere distance measurements pairwise for both types of distance as follows:
pick an arbi\-trary vertex $p$ and determine all $\delta$-spheres $S_p^\delta$ around it, up to $\delta\! =\! 15$. Check whether 
there is at least one vertex $p'\!\in\! S_p^{15}$ which is maximally timelike separated from $p$. If this is the case, randomly pick
both a maximally spacelike and a maximally timelike separated vertex $p'$ from each spherical shell (such vertices will always exist) and compute 
the average distance $\bar{d}(S_p^\delta,S_{p'}^\delta)$ for either, resulting in two sets of 15 measurement points each. 
If this is not the case, move to the next randomly chosen vertex $p$ and repeat the process.
By discarding the instances where (some or all of the) maximally timelike separated vertices do not exist one avoids an inhomogeneity in the 
comparative measurements. Since this happened only in about half of all cases, it seems reasonable to assume that the remaining
measurements are still representative of the overall geometry. 

\begin{figure}[t]
\begin{center}
\includegraphics[width=0.8\textwidth]{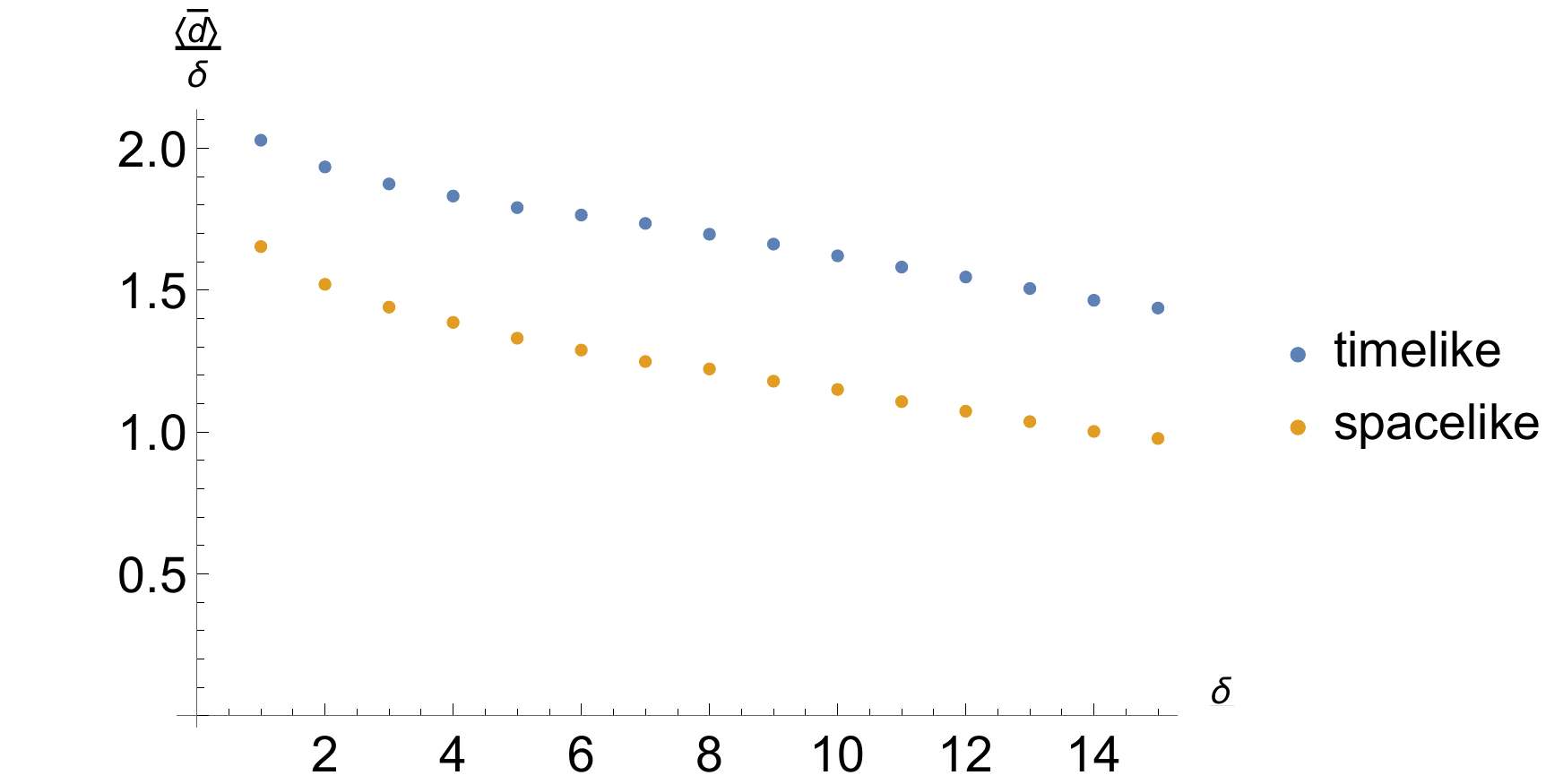}
\caption{Normalized average sphere distance $\langle \bar{d}\rangle /{\delta}$ as a function of the dual link distance $\delta$ at volume 
$N_{41}\! =\! 150k$, for sphere centres which are maximally timelike (blue) or maximally spacelike (yellow) separated. (Error bars are
smaller than dot size.)
}
\label{fig:cdtdir1}
\end{center}
\end{figure}
The results of our measurements of the expectation values $\langle \bar{d}\rangle /{\delta}$, performed at volume $N_{41}\! =\! 150k$, 
are shown in Fig.\ \ref{fig:cdtdir1}. Clearly, the measurements for spheres with maximally timelike displacement follow a
distinct trajectory from those with a maximally spacelike distance. They are systematically bigger than the latter, and appear to be
shifted vertically. What does this behaviour imply for the Ricci curvature? Recall from formula (\ref{qric}) that the quantum Ricci curvature
is related to the {\it deviation} of the sphere distance from a constant behaviour, with a positive deviation indicating negative curvature and
a negative deviation positive curvature. On continuum manifolds, the constant $c_q$ characterising flat behaviour depends only
on the dimension of the manifold \cite{qrc1}. In general, as remarked earlier, $c_q$ is a non-universal number depending on the
piecewise flat space under consideration. 

More relevant to our present discussion is the fact that the non-universality of $c_q$ also shows up as a discretization effect
on a single, given space when computing the directional dependence of the sphere distance. More specifically, when
investigating regular tilings of two-dimensional flat space we found a non-trivial dependence of $c_q$ on the 
direction of the difference vector between the sphere centres relative to the fixed lattice grid \cite{thesis}.
The presence of such an effect is not surprising in view of the properties of the link distance,
which is a crucial ingredient in the construction.
Using the link distance to measure lengths and distances on a regular square lattice, say, 
implies a well-known anisotropy when considering the set of vertices $p'$ at a distance $\delta$ from a given vertex $p$: 
whenever $p$ and $p'$ are separated by a straight sequence of lattice links, their Euclidean distance with respect to
the underlying flat space is given by $\delta\ell$, where $\ell$ is the length of a link. For the opposite extreme, when the
difference vector between $p$ and $p'$ lies along a lattice diagonal, the corresponding geodesic along lattice links is a 
zigzag or ``staircase" sequence of links, and the corresponding Euclidean distance $\delta\ell/\sqrt{2}$ between $p$ and $p'$ involves
a different proportionality factor. 

Since the quantum Ricci curvature $K_q(p,p')$ in (\ref{qric}) is related to a {\it quotient} of distances, it is not immediately
clear how it is affected by the anisotropic behaviour of the link distance. What we found on the two-dimensional square and
hexagonal lattices is that the only effect of non-straight, zigzag geodesics 
on the normalized average sphere distance $\bar{d}/\delta$ is a lowering of the
constant term $c_q$. The more the geodesic link distance between the two sphere centres overestimates their ``true"
geodesic distance, the smaller the value of $c_q$ becomes. 

\begin{figure}
\centerline{\scalebox{0.65}{\rotatebox{0}{\includegraphics{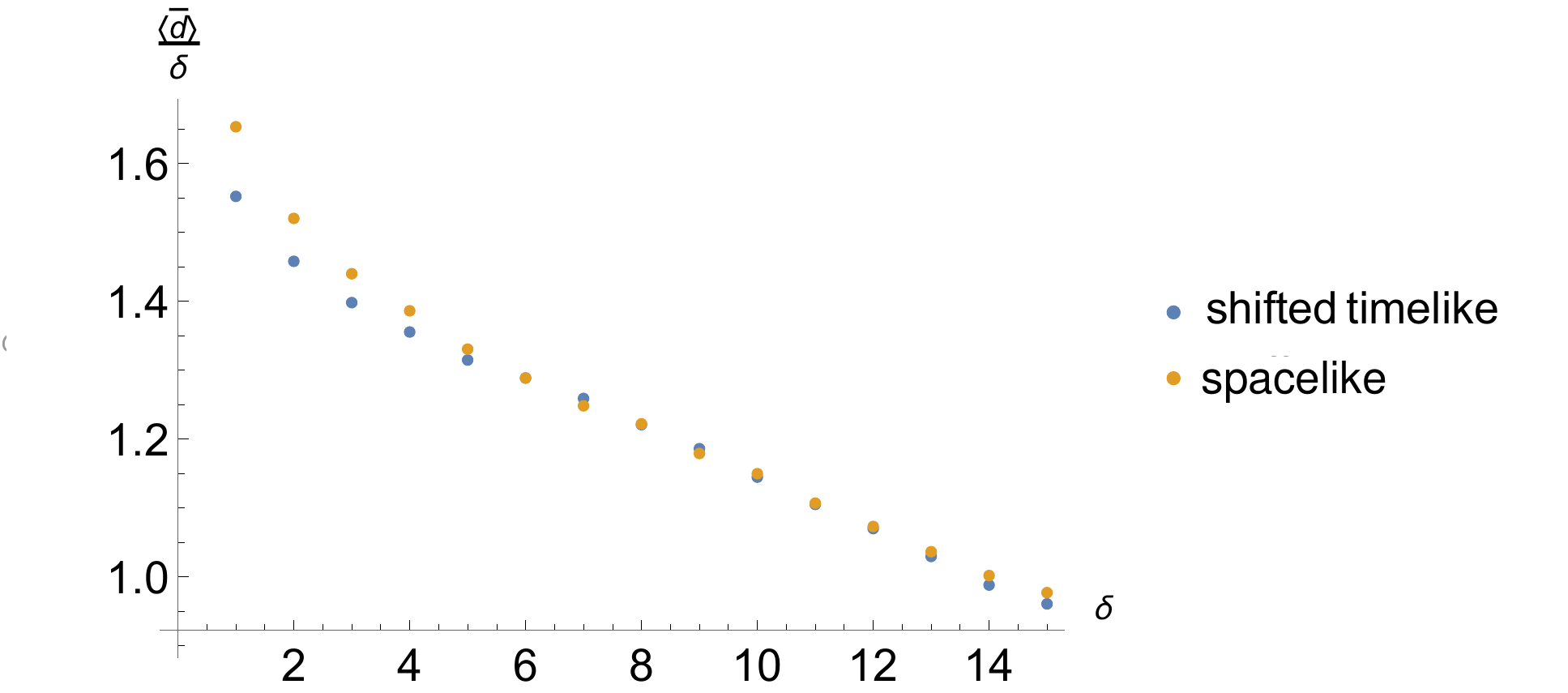}}}}
\caption{Normalized average sphere distance $\langle \bar{d}\rangle /{\delta}$ as a function of the dual link distance $\delta$ at 
volume $N_{41}\! =\! 150k$. Compared to Fig.\ \ref{fig:cdtdir1}, the data points for the maximally timelike separated spheres
have been shifted along the vertical axis by $-0.476$ to make the two data sets agree at $\delta\! =\! 6$.}
\label{fig:cdtdir2}
\end{figure}

While we certainly cannot claim to have studied this effect exhaustively, these findings provide a possible explanation for
why the sphere distance measurements in the time- and spacelike directions on the CDT configurations behave differently
and why this does not necessarily imply that the corresponding Ricci curvatures are different. The discrete lattice structure 
of the CDT configurations clearly has anisotropic features, although these are milder when working with the dual lattice
as already noted earlier. To examine this issue further we subtracted a constant offset from the maximally timelike
data depicted in Fig.\ \ref{fig:cdtdir1}, such that for $\delta\! =\! 6$ its value for $\langle \bar{d}\rangle/\delta$ coincides with 
that of the maximally spacelike data. As illustrated by Fig.\ \ref{fig:cdtdir2}, the only differences which then remain 
between the two data sets are for small $\delta$, which indicates that the short-distance artefacts affect time- and spacelike 
measurements differently.
More importantly, for the entire remaining range $\delta\! >\! 6$ the match is almost perfect. 
This suggests strongly that the observed difference is due entirely to a nonvanishing difference $\Delta c_q$ -- related
to an anisotropy of the underlying discrete lattice structure -- but that the quantum Ricci curvature behaves isotropically,
giving further support to the interpretation of the quantum universe in terms of a constantly curved de Sitter geometry.

\section{Summary and outlook}
\label{sec:concl}

We set out to understand the quasi-local geometric nature of the extended universe generated dynamically in
nonperturbative quantum gravity \`a la CDT. We asked to what extent it may resemble
a space of constant curvature and, more specifically, that of a four-sphere, a comparison which is suggested by the
excellent match of the expectation value of the global volume profile of the quantum universe with that of a classical de Sitter space. 
Until recently, we did not have suitable nonperturbative curvature observables allowing us to address this
more local aspect of the quantum geometry. This situation has changed with the introduction of the new quantum Ricci 
curvature \cite{qrc1}. After implementing and measuring this curvature in various well-controlled settings of classical piecewise flat
spaces and in a nonperturbative model of two-dimensional quantum gravity \cite{qrc1,qrc2}, with promising results, 
we have reported above on the first
implementation of the quantum Ricci curvature in the physically relevant case of four-dimensional nonperturbative quantum gravity. 

We would like to remind readers of the nontrivial nature of this undertaking, which did not come with any a priori guarantee 
of an interesting outcome. 
Firstly, due to the limitations of the numerical set-up there might have been no values of $\delta$ for
which the data are not strongly affected by either short- or long-distance lattice artefacts. This situation was illustrated by our
initial, inconclusive curvature measurements using the link distance. Secondly, in the limited scale range accessible when using
the dual link distance, no onset of semiclassical behaviour might have been found. While not per se uninteresting, this
could have had seve\-ral explanations: a dominance of pure quantum behaviour without any
classical features, an indication that the quantum Ricci curvature is not a useful observable in a Planckian regime, 
or a sign that the quantum universe has less in common with a classical de Sitter geometry after all. 

Instead, we found something much more interesting, which deserves to be called a new breakthrough result in CDT quantum gravity: 
over the range of quasi-local scales we could investigate,
the quantum universe appears to be posi\-ti\-vely curved, in a way that is compatible with the behaviour of a round four-sphere, 
a Euclidean de Sitter space! In addition to the sphere-like behaviour of the expectation value of the coarse-grained, spatially and
directionally averaged Ricci scalar,  
we also found evidence that the Ricci curvatures in space- and timelike directions are the same, within measuring accuracy. 
This is significant because the discrete lattice structure of the CDT configurations by construction has anisotropic features. 
Nevertheless, in terms of the direction-sensitive quantum Ricci curvature observable, local isotropy appears to be restored 
dynamically. This complements the well-known, earlier finding of an emergence of {\it global} isotropy in CDT's de Sitter phase: 
despite the fact that all configurations have topology $S^3\!\times\! S^1$, the ensemble is driven dynamically
towards a quantum geometry whose volume profile is that of a de Sitter space, with an effective $S^4$-topology \cite{desitter,cosmo}.    

It would be desirable to extend our investigation to larger volumes $N_{41}$, in order to further improve the
fits to the continuum curves and to allow for a finite-size scaling ana\-lysis of a similar quality as was possible
in two-dimensional quantum gravity \cite{qrc2}. This may also reveal quantum imprints in the behaviour of the average sphere
distance, which we are currently unable to resolve. 

Our measurements in four-dimensional quantum gravity 
underline the remarkable robustness of the quantum Ricci curvature, in the sense of displaying semiclassical behaviour
on length scales within an order of magnitude above the Planck scale. They also reconfirm the role of this quantity
as a useful new tool in our small, but growing arsenal of nonperturbative quantum gravity observables. It has allowed us to go one step
further in understanding the nature of the quantum geometry that emerges from the background-free, nonperturbative 
gravitational path integral, and is hopefully bringing us a step closer to understanding its phenomenological implications.

\subsection*{Acknowledgments} 
This work was partly supported by the research program
``Quantum gravity and the search for quantum spacetime" of the Foundation for Fundamental Research 
on Matter (FOM, now defunct), financially supported by the Netherlands Organisation for Scientific Research (NWO).

\end{document}